\documentclass[review]{elsarticle}

\usepackage{lineno,hyperref}
\modulolinenumbers[5]
\usepackage{amsmath,amsfonts}
\usepackage{algorithmic}
\usepackage{algorithm}
\usepackage{array}
\usepackage[caption=false,font=normalsize,labelfont=sf,textfont=sf]{subfig}
\usepackage{textcomp}
\usepackage{stfloats}
\usepackage{url}
\usepackage{verbatim}
\usepackage{graphicx}
\usepackage{tabularx,booktabs,textcomp}
\usepackage{amsthm}

\usepackage{amsmath,amssymb,amsthm}

\DeclareUnicodeCharacter {2061} {}

\begin{document}

\begin{frontmatter}

\title{Evaluating the Planning and Operational Resilience of Electrical Distribution Systems with Distributed Energy Resources using Complex Network Theory}

\author[1,2]{Divyanshi Dwivedi}

\author[2]{Pradeep Kumar Yemula}

\author[1]{Mayukha Pal\corref{mycorrespondingauthor}}
\ead{mayukha.pal@in.abb.com}
\cortext[mycorrespondingauthor]{Corresponding author}


\address[1]{ABB Ability Innovation Center, Asea Brown Boveri Company, Hyderabad, 500084, Telangana, India} 

\address[2]{Department of Electrical Engineering, Indian Institute of Technology, Hyderabad, 502205, Telangana, India}

\begin{abstract}

Electrical Distribution Systems (EDS) are extensively penetrated with Distributed Energy Resources (DERs) to cater the energy demands with the general perception that it enhances the system's resilience. However, integration of DERs may adversely affect the grid operation and affect the system resilience due to various factors like their intermittent availability, dynamics of weather conditions, non-linearity, complexity, number of malicious threats, and improved reliability requirements of consumers. This paper proposes a methodology to evaluate the planning and operational resilience of power distribution systems under extreme events and determines the withstand capability of the electrical network. The proposed framework is developed by effectively employing the complex network theory. Correlated networks for undesirable configurations are developed from the time-series data of active power monitored at nodes of the electrical network. For these correlated networks, compute the network parameters such as clustering coefficient, assortative coefficient, average degree, and power law exponent for the anticipation; and percolation threshold for the determination of the network’s withstand capability under extreme conditions. The proposed methodology is also suitable for identifying the hosting capacity of solar panels in the system while maintaining resilience under different unfavorable conditions and identifying the most critical nodes of the system that could drive the system into non-resilience. This framework is demonstrated on IEEE 123 node test feeder by generating active power time-series data for a variety of electrical conditions using the simulation software, GridLAB-D. The percolation threshold resulted as an effective metric for the determination of the planning and operational resilience of the power distribution system.

\end{abstract}

\end{frontmatter}

\section{Introduction}

In the recent decade, an enormous increase in the penetration level of Distributed Energy Sources (DERs) is perceptible due to the world’s shift towards clean energy and it is a general perception that their integration enhances the resilience of Electrical Distribution Systems (EDS) \cite{8967258}. Also, dynamic thermal rating (DTR) is introduced to improve the resilience of EDS by increasing capacity, managing overloads, detecting faults, enabling adaptive operation, and optimizing resource utilization \cite{DTR1, DTR2}. By leveraging real-time information and dynamic adjustments, DTR enhances the system's ability to withstand and recover from disturbances, ensuring reliable and efficient power delivery \cite{DTR3, DTR4, LAWAL2023119635}. With the emergence of large DERs and DTR in EDS, infrastructure becomes more complex, thus the real-time data-driven quantifiable measure of the resilience of the system is needed. 

The term ``resilience" is given in 2018 by the US federal energy regulatory commission as the capability to withstand, respond, adapt, and prevent situations such as disruptive events, man-made attacks, and severe technical faults \cite{Nationelectricity}. The resilience of the EDS has gained significant adhesion after severely affected storm Sandy in the year 2012 \cite{ALMEIDA20123}. However, blackouts are a common occurrence across the world. The majority of these outages are short-lived and affect only small geographical areas, resulting in minimal consequences for those without electricity access for a particular time duration. As there have been notable instances of significant blackouts with far-reaching impacts over the past two decades, spanning the entire globe \cite{su15031974}. The major blackouts that have taken place within the European interconnected system over the past 20 years such as the fire from a short circuit in an autotransformer in Greece on 8 February 2021 for 1 hour affected 1 million consumers \cite{fire}, extreme weather conditions in Poland on 8 April 2008 for 0.4 hours affected 24 million consumers \cite{en14248286}, power system failure in Italy on 28 September 2003 for 14 hours affected 60 million consumers \cite{UCTE}. These blackouts may affect the functioning of traffic control systems in major cities resulting in chaotic street conditions. Similarly, industrial operations will come to a halt, and there may be disruptions in telecommunication services. Situations such as failing artificial respirators in hospitals and individuals finding themselves trapped in elevators become particularly critical as vital electrical equipment stops supplying energy to the infrastructure \cite{app13010083}. Thus, it is imperative to conduct a resilience evaluation of EDS in real-time through a data-driven method to prevent blackouts, as the causes of them are frequent, highly complex and may even result in a complete system shutdown.

Some exiguous theoretical definitions and metrics had been proposed to practice resilience in electrical utilities \cite{MISHRA2021110201,8586495}. A resilient network can avoid discontinuity in supply to critical loads when exerted by undesired conditions such as power quality events, momentary interruptions, sustained outages, brownouts, and blackouts. Thus, the resilient system has the objective to maintain supply continuity and increase the time frame up to which continuity would be maintained. There are indices like the System Average Interruption Frequency Index (SAIFI) and System Average Interruption Duration Index (SAIDI) which help in computing the reliability of the system \cite{saidi}. However, they are not capable of anticipating the interruption and are not sufficient in identifying the withstand capability of the power distribution system in real-time. Over a few years for identifying the resilience of the power distribution network, the optimal partition technique has been used where the distribution network is partitioned into sub-networks. This technique exhibits a practice in which the initially local solutions are obtained from sub-networks and then the requisite information is communicated with neighbouring sub-networks to obtain an optimal global solution \cite{9458547}. An adaptive spectral graph partitioning algorithm is used which is based on node resettling and considers computational load balancing for synchronization, real power balance, and sub-networks resilience. It also ensures that the resilient power distribution network partitions can adjust to new operating conditions. Another approach considered for resilient distribution network analysis is the scheduling coefficient which aims to maximize the load recovery in minimum recovery time. It is achieved using a multi-objective improved simulated annealing algorithm which helps identify the optimal scheduling solution \cite{9041674}. Many papers have also been projected for resilient distribution networks for restoring critical loads \cite{r11,r22}. There are also other tools available such as the state estimation technique, which helps in studying the impact of rising penetration of DERs \cite{a11,a22}, but these approaches are unable to visualize the low observability condition present in the distribution system occurring because of heterogenous nature of data and measurements.

Measuring the resilience of power distribution systems requires the definition of its goals and metrics \cite{r1}. To assess the system's ability to withstand challenges, the authors proposed a metric based on the adaptive capacity of its assets \cite{r3}. Another metric examines the impact of short-term events on long-term operational resilience \cite{code}. A code-based metric was defined to quantify the resilience of the power distribution system and was proposed that utilizes the effect of short-term threats in predicting the long-term threat. However, this work did not anticipate the short-term threat to the system which could result in supply interruption. Additionally, a probabilistic metric is proposed to quantify the operational resilience of electric power distribution systems to high-impact low-probability (HILP) events \cite{r8}. Although these metrics evaluate the operational resilience of the power distribution system, they fail to consider the entire infrastructure as a unified observable quantity, thereby limiting the breadth and efficiency of the overall resilience perspective. Other resilience aspects are addressed by performance measures that account for the system's anticipation, endurance, and recovery capabilities in operational and planning contexts for EDS \cite{awr}.  The authors utilized graph theory, multi-criteria decision-making process, and component constraints for the metrics computation in this work. However, the work did not discuss and propose the optimal hosting capability of DERs while computing the resilience of EDS. Though it discussed the critical loads, their identification in the network is not achieved, which helps circumvent the interruption in the system’s operation. Furthermore, metrics based on the resourcefulness, rapidity, robustness, and adaptability of the system are proposed \cite{r5, r6}. However, these metrics fail to integrate the system's characteristics based on the progression of events, leading to an inadequate evaluation of resilience that impacts system operations. Another framework is proposed that utilizes geographic information systems (GIS) in predicting the risk levels of distribution networks \cite{GIS}. However, in this paper authors did not quantify the resilience of the system and the load-serving capability effectively. 

To address the above-mentioned issues, we propose the use of the percolation threshold, as a metric that follows the progression of events, involving end users and providing an effective evaluation of resilience. A percolation threshold is a statistical tool from the complex network theory that signifies the state transition of the system in response to extreme events, making it a suitable metric for energy resilience planning and operation. Recent studies suggest that complex network theory is suitable to be used for modelling the system, with the percolation threshold assessing the probability of failure for critical nodes and edges \cite{GNN}. Complex network analysis provides a better alternate statistical tool to understand the salient features of a network by visualizing the low observability conditions and giving its global perspective \cite{PAGANI20132688}. A complex network is used to identify and resolve the issues such as overloading in power systems, failures, and blackouts. This has been achieved by computing the parameters of complex networks and analyzing them to interpret critical nodes in the electrical network \cite{7381138}. Similarly, there has been a focus on identifying possible vulnerabilities, outages, and blackouts when new complexity gets introduced in the system to make it more reliable and secure \cite{9183667}. Centrality analysis technique from the complex network is reportedly used in finding the optimal placement of microgrids, as with the increase in DER penetration, the system is required to ensure its resilience, voltage stability, minimum power loss, and line loading \cite{a1}. Thus, in our proposed work, we have put forward an effective framework utilizing the complex network and machine learning to anticipate the occurrence of interruption threats and the system withstand capability. It also computes the hosting capability of DERs in the distribution system to enhance the system’s resilience.

In this paper, we propose a methodology in which complex network theory with non-linear dynamical parameters and machine learning algorithms are used to produce the metrics for system resilience. We considered the IEEE 123 node test feeder system for implementation of the proposed framework, simulated using GridLAB-D and generated the time-series data of active power at system nodes. Power supply continuity is a key aspect for the system to be resilient, thus we performed the analysis by taking active power supply and critical loads operations into consideration. For that steady state simulation is performed considering the load’s profile for 24 hours and then gradually solar panels were introduced in the system to check its effects and identify critical nodes. We also modified the standard IEEE 123 node test feeder’s loading and overhead line conductor parameters to explore system resilience under different conditions like the impact of weather, overloading, and unbalancing in the system. Our approach possibly provides a better resilient distribution system which would be more reliable and stable under normal and resilient conditions for varying DER penetration. We have also exploited the diverse visual methodology analysis to use complex networks for effectively visualizing the correlation between the generated data of the system with non-linearity arising under different analysis conditions. The key contributions of the proposed methodology are as follows:

\begin{enumerate}
    \item A complex network-based approach is proposed for evaluating the planning and operational resilience of power distribution systems under extreme events and determining the withstand capability of the electrical network.
    \item Correlated networks for EDS are obtained by taking active power supply and critical loads operations into consideration.
    \item Complex network parameters such as clustering coefficient, assortative coefficient, average degree and power law exponent are found suitable for the anticipation of system resilience; and percolation threshold for the determination of the network’s withstand capability under extreme conditions.
    \item Identification of hosting capacity of solar panels in the system while maintaining resilience under various unfavourable conditions and the most critical nodes of the system that could drive the system into non-resilience.
\end{enumerate}

This manuscript is organized with section \ref{section:MM} discussing materials and methods while section \ref{section:data} discusses the step-by-step procedure for implementation of the method on simulated data and its description. Section \ref{section:result} details the results and discusses the observed characteristics. Section \ref{section:conclusion} concludes the study with our inferences.

\section{Materials and Methods}
\label{section:MM}

\subsection{Electrical distribution system as complex network}

Consider an electrical distribution network (EDN) represented by graph $G=(V,E)$ where $V={\{1,2,…..N\}}$ is a set of nodes/vertices and $E$ is a subset of $V \times V$ that represents the edges $(m,n) \in V$ where $m \neq N$ \cite{NOVOSELNIK2015136}. There exists a swing node/generator node represented by $N$. $\Bar{V}={\{1,2, \dots, N-1\}}$  which denotes all the system nodes except the swing node. In such a system, we place photovoltaic (PV) panels at all the nodes except the swing node. We can define the system model as:
\begin{itemize}
    \item $PV \subseteq \Bar{V}$, a set of nodes having PV panels.
    \item $LD \subseteq \Bar{V}$, a set of nodes having loads connected. 
\end{itemize}
 
Considering a $node-(m) \in \bar{V}$ of the electrical network, the real and reactive power at every time instant $t$ is given by:
\begin{equation}
    P_{m,t}^i= P_{m,t}^{PV} - P_{m,t}^{LD}=\sum_{j\in \bar{V}} \delta_{mn,t}  P_{mn,t}         \hspace{0.5cm} \forall \in \bar{V}
\end{equation}
\begin{equation}
    Q_{m,t}^i= Q_{m,t}^{PV} - Q_{m,t}^{LD}=\sum_{j\in \bar{V}} \delta_{mn,t}  Q_{mn,t}         \hspace{0.5cm} \forall \in \bar{V}
\end{equation}

where, \begin{equation}
    P_{mn,t}= g_{mn} V_{m,t}^2 - V_{m,t} V_{n,t} (g_{mn} cos⁡\theta_{mn,t}+b_{mn} sin\theta_{mn,t})	
\end{equation}
\begin{equation}
    Q_{mn,t}= V_{m,t} V_{n,t} (g_{mn} cos⁡\theta_{mn,t}+b_{mn} sin\theta_{mn,t})- b_{mn} V_{m,t}^2	
\end{equation}

Here, $P_{m,t}^{LD} $ and $Q_{m,t}^{LD}$ are the real and reactive power of load connected to $node-m \in LD$ at time $t$ respectively. $P_{m,t}^{PV} $ and $Q_{m,t}^{PV}$ are the real and reactive power of PV generation to node $m \in PV$ at time $t$ respectively. $V_{m,t}$ is the voltage magnitude at $node-m \in V$ at time $t$. $\theta_{mn,t}$ is the voltage angle difference between nodes $(m,n) \in V$  at time $t$. $P_{mn,t}$ and  $Q_{mn,t}$ are the real and reactive power transferred from $node-m$ to the network through line $(m,n) \in E$ at any time instant $t$. $\delta_{mn,t}$ is operation status of line $(m, n) \in E$ at any time instant $t$. For  $\delta_{mn,t}=1$, the lines are operational, $\delta_{mn,t}=0$ means lines are not operational. $z_{mn}$ denotes impedance of line, and  $y_{mn}=[z_{mn}]^{-1}$ as its admittance. Thus, $y_{mn}=g_{mn}+j b_{mn}$, where $g_{mn}$ is conductance and $b_{mn}$ is susceptance of line $(m, n) \in E$ at time $t$.

The voltage magnitude at node $m \in V$  lies within lower and upper bounds as: 
\begin{equation}
    V_{lb} \leq V_{m,t} \leq V_{ub} \hspace{0.5cm} \forall \in {V}
\end{equation}

PV panel generated output $P_{m,t}^{PV}$ is intermittent as it depends on meteorological conditions. However, let us consider the presence of solar irradiance, then the PV real power produced is modelled as \cite{NOVOSELNIK2015136}:
\begin{equation}
    P_{m,t}^{PV}= \alpha_{m,1} S_{m,t}+ \alpha_{m,2} T_{m,t}+\alpha_{m,3} S_{m,t} T_{m,t}  \hspace{0.5cm} \forall \in {PV} 
\end{equation}

where, $T_{m,t}$  is the temperature and $S_{m,t}$ is the incident solar irradiance at $node-m \in PV$ at time $t$. $\alpha_{m,1}$, $\alpha_{m,2}$, and $\alpha_{m,3}$ are the PV model parameters. The reactive power $Q_{m,t}^{PV}$ depends on the power electronics equipment present and we can represent reactive power injections using lower and upper bounds as:
\begin{equation}
    Q_{lb} \leq Q_{m,t} \leq Q_{ub} \hspace{0.5cm} \forall \in {PV}
\end{equation}

\subsection{Resilience in electrical distribution system}

\begin{figure}
  \centering
  \includegraphics[width=3.2in]{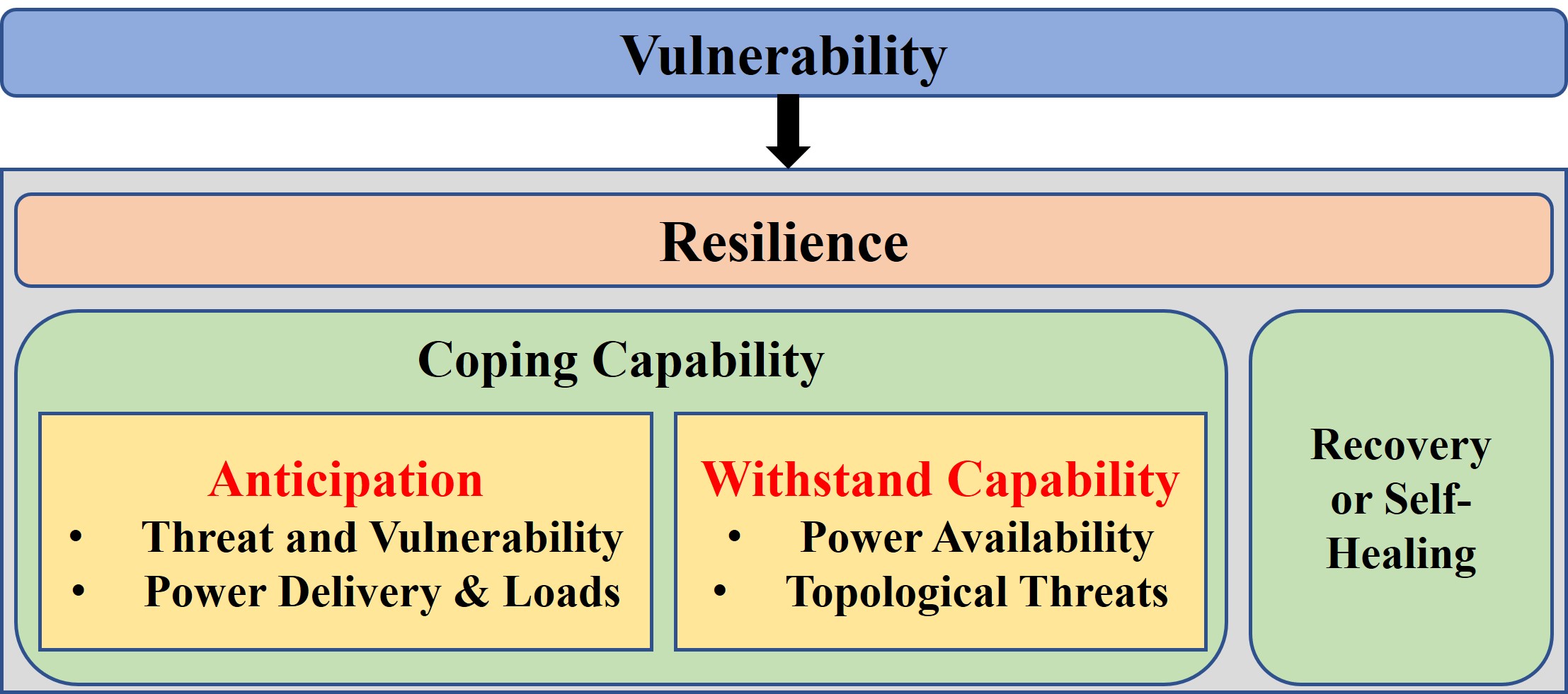}\\
  \caption{Aspects of understanding resilience in EDS.}
  \label{fig:resilience}
\end{figure}

With high penetration of DERs in EDS, the system becomes more reliable and resilient, but DER also pertains stochastic nature which could affect the distribution grid operation. Altogether with unwanted events, these DERs would severely affect the grid performance and thus EDS must have the capability to cope with the changes effectively, which is termed as short-term resilience of the system \cite{cscc}. Resilience is a sub-category of vulnerability that has two aspects to investigate, first coping capability and second recovering as depicted in Figure \ref{fig:resilience}. The resilient system should withstand the circumstances arising but in case fails to do so then must have the capability to recover itself \cite{scc}. In this work, we have evaluated the coping capability of an EDS when incorporated with solar PV panels under varying circumstances.

For analyzing the resilience, we used complex network theory as a computational tool. The concept of percolation theory was applied in this study for analyzing the system resilience of a network modelled using a complex network framework and helps in identifying the critical nodes of the system \cite{iscc}. Percolation theory seems to be successful as a qualitative guide to the resilience of networks \cite{book}. Percolation theory helps in identifying the ‘phase-transition’ when nodes/edges are removed in a network, thus effectively utilized as a statistical tool for identifying the operational transition occurring in EDS.

\subsection{Percolation threshold in identifying resilience}

During normal operation, having all the operating conditions within their specified limits, the probability of operational nodes is $\rho=1$. If a system is introduced with undesired events, then it will affect the nodes of the system, although sometime nodes will remain operational and have a probability less than $1$ $(\rho <1)$ or it can become non-operational with $(1 -\rho)$ as a probability. Thus, the threshold value for the probability under these events influences the operational nodes and is known as percolation threshold $\rho_c$, which identifies the \% of critical nodes in the system that could easily damage or break the network under undesired circumstances. When $\rho > \rho_c$, then the system is considered as resilient and for $\rho \leq \rho_c$, the system is considered weak that would not cope with the arising conditions. Percolation theory is found to be effective in obtaining both quantitative and qualitative measures of the resilience of networks \cite{PhysRevE.91.010801}. Computation of the percolation threshold is achieved by calculating the percolation strength of the network, followed by calculating susceptibility and the best estimate of the percolation threshold $\rho_c$ as the value of $\rho$ is where the susceptibility reaches its maximum \cite{assortative}. 
Percolation strength, 
\begin{equation}
    P_\infty(p)=  \frac{1}{NQ} \sum_{q=1}^Q [S_q(p)]
\end{equation}
where, $S(p)$ is the function of bond occupation probability, $p=e/E$, with $E$ the total number of edges, $e$ is the number of edges removed from the initial configuration and $N$ is the number of nodes. We repeat the entire process up to $Q$ times and compute the susceptibility.
\begin{equation}
 \chi(p)= \frac{\frac{1}{N^2 Q} [\sum_{q=1}^Q S_q (p) S_q (p) - P_\infty(p)]^2}{P_\infty (p)}
 \end{equation}
Percolation Threshold,
 \begin{equation}
 \rho_c= arg\{max_p \chi(p)\}			
 \end{equation}

There exist many percolations process with the most used types being bond percolation, explosive percolation, and site percolation which depends on how one considers the analysis in a lattice structure. Site percolation gives a view of the lattice as a rectangular array of squares whereas bond percolation gives a view of the graph as horizontal and vertical edges. For this study, we have considered bond percolation as it gives an analogy of how effectively and strongly nodes are connected through each other.

\subsection{Percolation in correlated networks}

Real-world networks are generally correlated networks. For the distribution system, a correlated network means how the time-series loading pattern at $node-m$ is correlated with the loading pattern at $node-n$. Similarly, with PV incorporation, it suggests how the time-series of the PV generation pattern at $node-m$ is correlated with the PV generation pattern at $node-n$. Usually, correlation in networks helps in attributing complex structures \cite{corr1,corr2}. Complex network derived out of correlation matrix also provides a great inference, especially with explosive percolation \cite{exp}. Time-series data when analyzed using a correlation matrix provides more robust and reliable solutions even though the data is noisy and ill-framed. To reduce the noises in data, thresholding on the correlated network may help in generating robust information from the networks. We verify that the explosive percolation for a correlated network provides similar inference as bond percolation hence, in our work we have implemented bond percolation on correlation networks. The methodology used for constructing a correlation network is explained in the next section.

\subsection{Complex network parameters for analyzing correlated networks}

The correlation matrix represents the relational information of multiple time-series data \cite{HAN2021107377}. The Pearson correlation coefficient helps identify the strength of the relationship between the networks. When the correlation coefficient of two time-series data having n number of nodes is calculated, we obtain the $n \times n$  coefficient matrix that provides relational information among the nodes. The value of these coefficients varies between $-1$ and $1$. The Pearson correlation coefficient of two time-series at nodes $m$ and $n$ is given by:
\begin{equation}
    PC= \frac{\sum [P_m^i (t)-\Bar{P_m^i}] [P_n^i (t)-\bar{P_n^i}]}{\sqrt{\sum [P_m^i (t)-\Bar{P_m^i}]^2 [P_n^i (t)-\bar{P_n^i}]^2}}
\end{equation}

where, $\bar{P_m^i}$  and $\bar{P_n^i}$ are the mean for the active power time-series of $P_m^i$ and $P_n^i$  respectively. This obtained correlation matrix plots the complete graph with all possible edges. To understand connection density among nodes of significance, the use of thresholding produces a sparse adjacency matrix that generates a desirable graph for the system from the correlation matrix. Correlation networks using a variety of such techniques help in removing noise from the data while analyzing the system under study \cite{NOVOSELNIK2015136}. In this work, we correlated the base configuration of EDS with a variety of conditions incorporating PV to generate the correlation matrix and then the complex network to compute various network parameters for electrical inferences. The network parameters are computed as given below:

\begin{enumerate}
    \item Average Degree (AD): For an undirected graph, it is defined as the average number of edges per node. Consider $V$ as several nodes/vertices and $E$ as the number of links/edges in the network, it is denoted by $\bar{K_i}$ and written as:
    \begin{equation}
        \bar{K_i}= \frac{2E}{V}
    \end{equation}
    It infers how well the network is connected, where a high value of average degree means the system is densely connected.
    \item Clustering Coefficient (CC): It gives the degree to which neighbours of a given node link to each other. Mathematically, we write it as:
    \begin{equation}
        C_i= \frac{(2L_i)}{K_i (K_i-1)}
    \end{equation}
    where $K_i$ is the degree of node $i$ and $L_i$ is the neighbour links of $node-i$. It is the coefficient that suggests how a graph tends to cluster together. High values of the clustering coefficient indicate the system is strongly connected.
    \item Minimum Degree (MD): It is the least degree of a node existing in the network representing the connectedness of the network.
    \item Assortative Coefficient (AC): It measures the level of homophyly of the graph. It has a value ranging from $-1$ and $1$, where for $r = 1$, the network is considered as a perfect assortative while for $r = 0$ the network is non-assortative and for $r =-1 $ the network is completely disassortative. When the value is $1$, it signifies strong nodes tend to connect with strong nodes and weak nodes tend to connect with weak nodes whereas, for value $-1$, it signifies that strong nodes tend to connect with weak nodes or vice-versa. The assortative coefficient is mathematically expressed as \cite{assor}:
    \begin{equation}
     r=\frac{L^{-1} \sum_i j_i k_i-[L^{-1} \frac{1}{2} \sum_i (j_i+k_i)]^2}{L^{-1} \sum_i j_i^2+k_i^2-[L^{-1} \frac{1}{2}\sum_i (j_i+k_i)]^2 }
    \end{equation}
    where, $j_i$, and $k_i$ are the degrees of the vertices at the ends of the $i^{th}$ edges, with $i=1,2, \dots, L$.
    \item Power Law Fit (PLF): It shows the relationship between two operating conditions of the system and indicates how the system at one operating condition varies as a power of another operating condition without concern about the system size. Mathematically, a power law is expressed as:
    \begin{equation}
        f(x)=ax^{-k}
    \end{equation}
\end{enumerate}

In this paper, we have identified how system parameters are relatively changing with the incorporation of PV panels and undesirable conditions. For a complex network, power-law degree distribution comes into existence only when the probability distribution of degree in any system follows the power law and then only the system is considered as resilient. These computed network parameters along with the percolation threshold are used for our analysis.

\section{Data and its Processing}
\label{section:data}

In this work, we first generated the time-series data for a standard IEEE 123 node test feeder using GridLAB-D, an open-source software that provides a platform to easily design a distribution system with the incorporation of DERs. GridLAB-D is a distribution-level power system simulator where simulation could be achieved either with event-driven mode or through sub-second simulation mode (delta mode). The reliability of GridLAB-D is well studied in EDS simulation, as it follows an agent-based simulation paradigm and the output obtained closely resembles the data collected from the smart grid demonstration project \cite{gridlab}. We simulated data using the event-driven mode, considering the system to be steady and have consistent and coherent characteristics. 

\begin{table}
\centering
\caption{IEEE 123 node test feeder system parameters}
\begin{tabular}{>{\centering}m{14em} c c}
    \toprule
    \multicolumn{1}{>{\centering}m{14em}}{System Parameters}  &
    \multicolumn{1}{>{\centering}m{14em}}{Values}\\
\midrule
Input Real Power	& 3620.498 kW\\
Input Reactive Power &	1324.365 kVAR\\
Input Power Factor &	0.939\\
Peak Real Power Demand &	3524.557 kW\\
Load Reactive Power Demand &	1940.830 kVAR\\
Solar PV Panel Rating	 & 30 kWp (at the single node)\\
\bottomrule
\end{tabular}
\label{tab:System_parameters}
\end{table}

\begin{figure}
  \centering
  \includegraphics[width=4in]{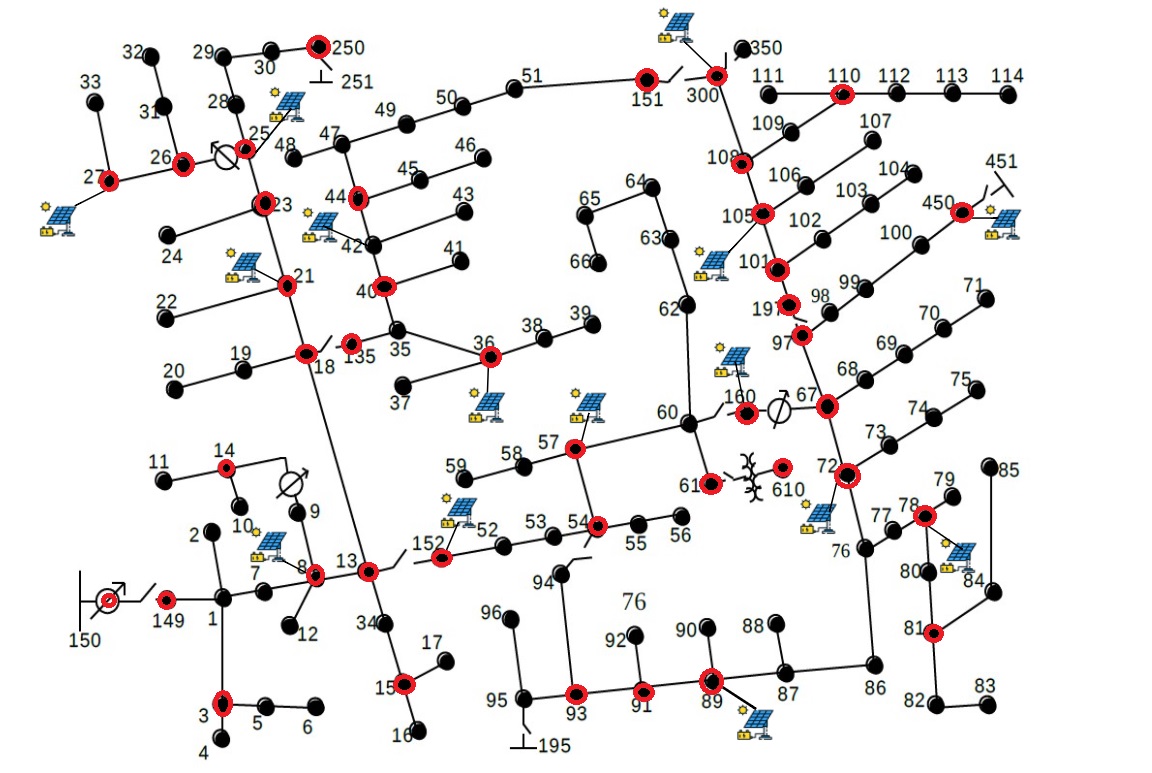}
  \caption{Single line diagram of IEEE 123 node test feeder with 40 meters nodes marked red with 40\% of meter nodes (i.e., 16 meters) incorporated with PV panels.}
  \label{fig:IEEE123}
\end{figure}

IEEE 123 node test feeder operates at a nominal voltage of 4.16 kV and comprises overhead as well as underground lines, loaded with constant current, impedance, and power with four regulators, capacitor banks, and switches. This system has all the components of a realistic distribution system, thus considered reliable for performing the data-driven analysis; system details are tabulated in Table \ref{tab:System_parameters}. IEEE 123 node test feeder consists of 85 constant loads, and we collected incoming real power on 40 nodes of the system which are marked red in Figure \ref{fig:IEEE123}. These nodes are referred to as Meter nodes, where we are calculating the incoming real power and other electrical parameters at these specific nodes considering the defined system characteristics. The generated time-series data includes various electrical parameters of the system such as real power and reactive power demand at loads, voltage, and currents at the nodes of the system. For further analysis, we have considered the real power demand of the loads and the generated power supply into consideration in our framework because the system’s resilience is typically modelled with load demand being served and real power supply continuity is maintained.

Non-linearity in the standard system is introduced with the incorporation of Solar PV panels in the system \cite{q35}. We first placed PV on 20\% (i.e. 8 nodes) out of 40 meters, then gradually increased the PV percentage to 40\% (i.e. 16-meter nodes), then 60\% (i.e. 24-meter nodes), then 80\% (i.e. 32-meter nodes) and finally 100\% i.e., placing PV at all meter nodes to analyze the impact and identify PV hosting capacity of the system while maintaining the resilience of the network. Here, meter node 150 is considered a swing node or generator node.

Energy resilient systems could sustain any circumstances arising out of nonlinearity introduction hence we have considered the following cases in our study:
\begin{itemize}
    \item Case- I: Normal Operating Conditions - Standard IEEE 123 node test feeder system. This case is denoted by $P$. 
    \item Case- II: High Consumption - Increase loading while maintaining system balancing to check system behaviour in overloading conditions. This case is denoted by $HC$. This case would help in identifying the resilience in terms of the power generation capability of the system to fulfil increasing load demands in the system.
    \item Case- III: High Resistance - Resistance increases for the overhead line conductor from 0.036 ohms to 2 ohms. This case is denoted by $HR$. This case is implemented to check the topological threats capability that the system could withstand.
    \item Case- IV: Imbalanced - Load values of PQ type loads increase with load imperfection which causes unbalance in the three-phase voltage. This case is denoted by $IB$. This case would analyze the impact of changes in the load operating conditions on the system’s resilience. 
\end{itemize}

Here, we are considering time-series data of real power at the meter nodes. To understand system behavior after incorporation of PV under these scenarios, we computed Pearson correlation coefficient matrices of size $40 \times 40$ from the generated real power time-series data as it helps in scrutinizing the electrical system performance based on complex network theory parameters and suggest the system node’s correlation when these events take place. We further obtained the complex correlated network that gives an analogy of how densely the system is interconnected. In this work, the correlated network is the measurement of similarity in the dynamics between the system with and without PV and is computed from the considered two time-series data. We considered the positively correlated network which depicts how two systems in different circumstances behave in a particular direction. We analyzed how to load power increases when the percentage of PV increase in the system, hence introduced a threshold value $T$ (i.e., we considered values greater than 0) and transformed the correlation matrix to an adjacency matrix by substituting all the values above the threshold $T$ as 1 otherwise 0. It generates a sparse matrix and reduces the complexity of understanding the correlation in the network. Furthermore, the results are consistent for different values of correlation thresholds thus we can state that our obtained networks are robust and reliable.

Further, various network parameters were computed from the obtained graphs including parameters such as average degree, correlation coefficient, assortative coefficient, minimum degree and power law exponent value for anticipation of threats. The percolation threshold is computed for checking the system’s withstand capability the threats. We computed bond percolation in the correlation network as it results in similar inference as observed in the case of explosive percolation. Bond percolation values were computed with 1000 iterations. 

Our analysis also finds out the optimal allocation topology for an example case of 40\% PV incorporated IEEE 123 node test feeder network to understand critical loads in the system. For this analysis, we have considered 15 combinations as mentioned in Table \ref{tab:Combo}, and then computed the percolation threshold following the process detailed in Figure \ref{fig:flow}. 

\begin{figure}
  \centering
  \includegraphics[width=4.6in]{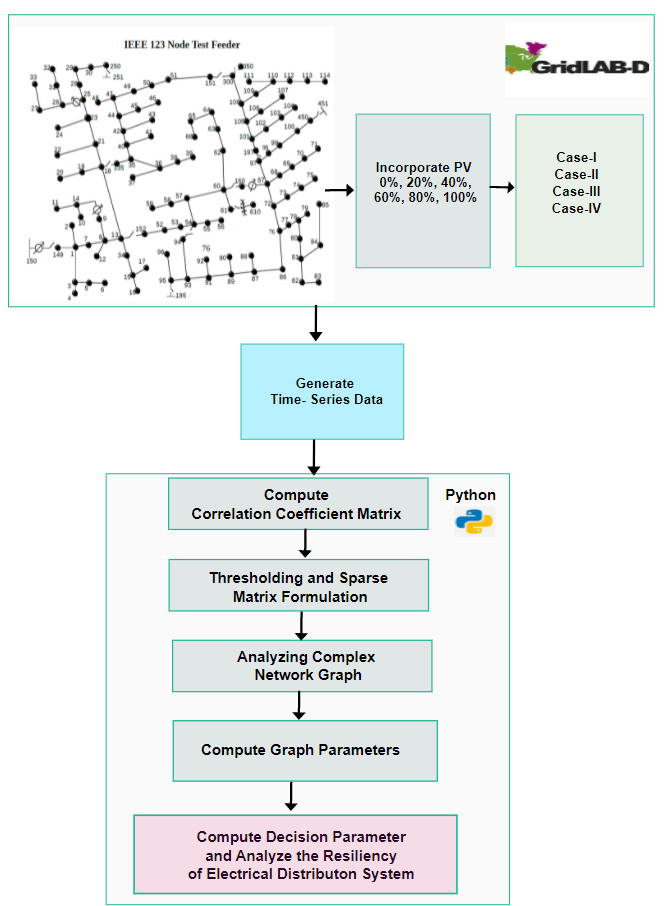}
  \caption{Proposed Methodology}
  \label{fig:flow}
\end{figure}

Percolation threshold as a parameter gives inference for the resilient distribution system and optimal topology with the incorporation of PV in the system including the critical load. We compute network parameters for each combination which was further subjected to feature selection using a random forest regressor machine learning algorithm to understand the network parameters that are highly sensitive to percolation threshold and hence to the system resilience, thereby identifying the indicator of the weak parameters in the system resilience. Of the various machine learning algorithms used for feature selection, random forest performed better with higher accuracy. 

\begin{figure}
  \centering
  \includegraphics[width=4.0in]{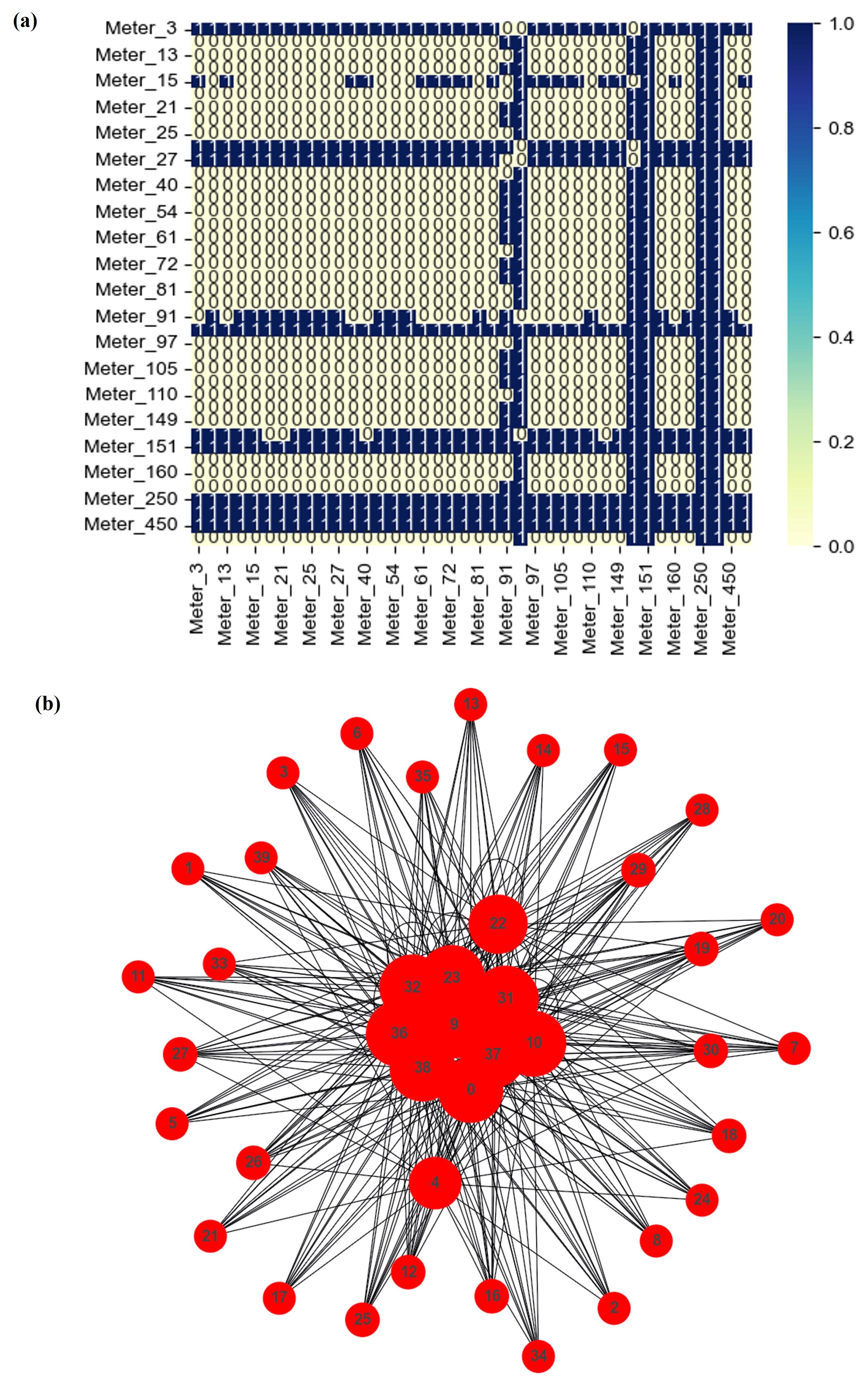}
  \caption{Positively correlated network obtained incorporating 0\% and 40\% of PV panels under
normal operating conditions; (a) the adjacency matrix for the correlation network, (b) the
correlation network.}
  \label{fig:flow}
\end{figure}

\begin{table}[]
\caption{Considered combinations for placing PV panels at 40\% of the meter nodes}
\begin{tabular}{>{}p{1.2cm}>{}p{0.25cm}>{}p{0.25cm}>{}p{0.25cm}>{}p{0.25cm}>{}p{0.25cm}>{}p{0.25cm}>{}p{0.25cm}>{}p{0.25cm}>{}p{0.25cm}>{}p{0.25cm}>{}p{0.25cm}>{}p{0.25cm}>{}p{0.25cm}>{}p{0.25cm}>{}p{0.25cm}>{}p{0.25cm}}
\toprule
\multicolumn{1}{>{}p{1.2cm}} {Combi- nation}  & \multicolumn{16}{>{}p{10cm}}{Node combinations for the 40\% meter nodes  incorporated with PV panels} \\
\midrule
C1  & 8   & 21 & 25 & 27 & 36 & 44 & 57  & 78  & 89  & 97  & 105 & 152 & 160 & 197 & 300 & 450 \\
C2  & 3   & 8  & 13 & 14 & 15 & 18 & 21  & 23  & 25  & 54  & 57  & 61  & 93  & 27  & 149 & 152 \\
C3  & 3   & 8  & 13 & 14 & 15 & 23 & 25  & 26  & 27  & 54  & 93  & 18  & 21  & 250 & 149 & 152 \\
C4  & 3   & 8  & 13 & 14 & 15 & 23 & 25  & 26  & 250 & 54  & 93  & 18  & 21  & 61  & 149 & 152 \\
C5  & 13  & 15 & 18 & 21 & 23 & 25 & 40  & 44  & 54  & 57  & 89  & 91  & 93  & 197 & 135 & 152 \\
C6  & 3   & 23 & 27 & 36 & 44 & 57 & 67  & 81  & 89  & 97  & 135 & 197 & 250 & 18  & 300 & 450 \\
C7  & 151 & 67 & 72 & 78 & 81 & 97 & 101 & 105 & 108 & 110 & 151 & 160 & 197 & 250 & 300 & 450 \\
C8  & 18  & 21 & 23 & 25 & 26 & 27 & 36  & 40  & 44  & 57  & 61  & 135 & 160 & 300 & 250 & 610 \\
C9  & 3   & 8  & 13 & 14 & 15 & 54 & 57  & 87  & 89  & 91  & 93  & 149 & 152 & 450 & 160 & 197 \\
C10 & 3   & 8  & 13 & 14 & 15 & 18 & 21  & 23  & 25  & 26  & 27  & 149 & 250 & 197 & 450 & 300 \\
C11 & 3   & 8  & 13 & 14 & 15 & 18 & 21  & 23  & 25  & 36  & 40  & 44  & 135 & 450 & 151 & 250 \\
C12 & 3   & 8  & 13 & 14 & 15 & 54 & 57  & 61  & 89  & 91  & 93  & 152 & 160 & 450 & 195 & 610 \\
C13 & 18  & 21 & 23 & 25 & 26 & 27 & 67  & 72  & 78  & 81  & 97  & 250 & 450 & 450 & 300 & 160 \\
C14 & 3   & 8  & 13 & 14 & 15 & 54 & 57  & 101 & 105 & 108 & 110 & 149 & 152 & 450 & 197 & 300 \\
C15 & 36  & 40 & 44 & 54 & 57 & 61 & 67  & 72  & 97  & 135 & 151 & 152 & 160 & 18  & 450 & 610 \\
\bottomrule
\label{tab:Combo}
\end{tabular}
\end{table}

\section{Results and Discussion}
\label{section:result}
IEEE 123 node test feeder is considered for implementing the proposed methodology and further analysis of resilience with the incorporation of PV panels under undesirable circumstances. Time-series data for the system is generated using GridLAB-D and the active power of system nodes is taken into consideration for analysis. 

\subsection{IEEE 123 node test feeder with the incorporation of PV panels for various analyzed cases}

For normal operating conditions i.e., Case-I, with increasing incorporation of PV panels at meter nodes at 20\%, 40\%, 60\%, 80\% and 100\% of 40 meter nodes; the correlation network is obtained between the standard system and PV incorporated system. The percolation threshold is the deciding parameter that infers the transition of the system from resilient to non-resilient. The lower value of the percolation threshold is desired when the transition in systems’ operation is taken into account \cite{RAHAMAN2022107,RAMANUJAM20171}. On the other hand, we expect our system to be resilient when incorporating more PV panels, thereby the operating performance of the system should not get affected by the changes occurring in the system. For a resilient EDS, it is desired to have a high value of percolation threshold, which means the system will not face any transition easily when encountering the changes, rather if the percolation threshold is low it suggests that the system will undergo the transition to non-resilient easily and may not persist the changes \cite{iscc}. It is worth observing that computed percolation thresholds for correlated networks are resulting low values as reported in Table \ref{tab:nomenclature} for our analysis because the general EDS is highly correlated and densely clustered thus macroscopic clusters form at low values of occupation probability \cite{PhysRevE.91.010801}. Table \ref{tab:nomenclature} depicts the change in systems’ resilience in terms of percolation threshold which suggests such undesired events are hampering the system's resilience. In Table \ref{tab:nomenclature} the nomenclature used for naming the graph like $P_{0\%}-HC_{0\%}$ suggests that we are finding a correlation between Case-I denoted by P with 0\% PV incorporation and Case-II denoted by HC with 0\% PV incorporation. This naming convention was used for all the network graphs for different cases under different PV incorporation.

\begin{table}
\centering
\caption{Sample representation of Percolation threshold for the networks under the influence of undesired events}
\begin{tabular}{>{\centering}m{6.5em} c c}
    \toprule
    \multicolumn{1}{>{\centering}m{6.5em}}{Graph Name}  &
    \multicolumn{1}{>{\centering}m{6.5em}}{Percolation Threshold}\\
    \midrule
    $P_{0\%} - HC_{0\%}$ & 0.0346 \\
    $P_{0\%} - HR_{0\%}$ & 0.0342 \\
    $P_{0\%} - IB_{0\%}$ & 0.0339   \\ 
\bottomrule
\end{tabular}
\label{tab:nomenclature}
\end{table}

With various case studies, we analyze and discuss in detail the circumstances where the system is experiencing the transition and may not sustain the changes that occurred for the considered cases. Table \ref{tab:percolation_main} suggests the distribution system resilience for various undesirable events with the integration of PV. 

\subsubsection{Discussion for Case-I:}
From Table \ref{tab:percolation_main} we could observe that the clustering coefficients for all the networks are above 80\% which depicts that the system is strongly connected and if any node gets interrupted, continuity of supply will be maintained as these are well-connected networks. The assortativity coefficient values of the networks are $\approx -0.70$, hence the networks are disassortative having their weak nodes establishing good relationships with the strong nodes. The significance of disassortative in an electrical network we explain with an example: when the overloading condition occurs at any node whereas other nodes have abundance power then these nodes could support the overloaded node to fulfil the demand without resulting in any break out in the system. In the analogy of the scenario in a complex network, those overloaded nodes are weak whereas the node with abundant power is strong. With a strong relationship between both nodes, the electrical system becomes stronger as it helps maintain supply continuity, resolving unwanted failures. Thus, in this case, the networks are strongly connected and withstand the incorporation of PV panels in the system. We could also observe that these networks follow power law as detailed in section \ref{section:MM}. 

\subsubsection{Discussion for Case-II:}
In this case, the system persistence is evaluated when load consumption increases. From Table \ref{tab:percolation_main}, the clustering coefficients of these networks are $\approx 0.75$ and compared to Case-I it is less resilient however the system is strong enough to cope with the changes. Assortativity coefficient values are also $\approx 0.67$ hence the networks are disassortative. Network parameters suggest the system here is not as strong as in the case of Case-I. 

\begin{table}
\caption{Network parameters for different cases with the incorporation of PV}
\begin{tabular}{cccccccc}
\toprule
Cases & Graph  &	\multicolumn{1}{>{}p{1cm}}{AD} &	\multicolumn{1}{>{}p{1cm}}{CC} &	\multicolumn{1}{>{}p{1cm}}{MD}	& \multicolumn{1}{>{}p{1cm}}{AC}	& \multicolumn{1}{>{}p{1cm}}{PLF} & \multicolumn{1}{>{}p{1cm}}{PT} \\
\midrule

Case-I   & $P_{0\%}-P_{20\%}$    & 18.15 & 0.83 & 10.00 & -0.70 & 3.55  & 0.052 \\
                          & $P_{0\%}-P_{40\%}$    & 17.75 & 0.85 & 10.00 & -0.69 & 3.75  & 0.054 \\
                          & $P_{0\%}-P_{60\%}$    & 17.70 & 0.84 & 10.00 & -0.69 & 3.75  & 0.054 \\
                          & $P_{0\%}-P_{80\%}$    & 17.70 & 0.85 & 9.00  & -0.70 & 3.15  & 0.056 \\
                          & $P_{0\%}-P_{100\%}$   & 17.65 & 0.85 & 10.00 & -0.69 & 3.77  & 0.057 \\
Case-II  & $HC_{0\%}-HC_{20\%}$  & 20.90 & 0.74 & 12.00 & -0.67 & 3.31  & 0.045 \\
                          & $HC_{0\%}-HC_{40\%}$  & 20.90 & 0.79 & 12.00 & -0.67 & 3.51  & 0.048 \\
                          & $HC_{0\%}-HC_{60\%}$  & 20.40 & 0.77 & 11.00 & -0.67 & 3.18  & 0.049 \\
                          & $HC_{0\%}-HC_{80\%}$  & 20.65 & 0.78 & 11.00 & -0.67 & 3.11  & 0.051 \\
                          & $HC_{0\%}-HC_{100\%}$ & 20.60 & 0.77 & 12.00 & -0.68 & 3.61  & 0.051 \\
Case-III & $HR_{0\%}-HR_{20\%}$  & 30.05 & 0.90 & 19.00 & 0.03  & 11.32 & 0.043 \\
                          & $HR_{0\%}-HR_{40\%}$  & 28.35 & 0.89 & 19.00 & 0.04  & 10.33 & 0.042 \\
                          &  $HR_{0\%}-HR_{60\%}$  & 27.55 & 0.80 & 18.00 & -0.53 & 3.86  & 0.040 \\
                          &  $HR_{0\%}-HR_{80\%}$  & 30.25 & 0.91 & 19.00 & -0.05 & 8.564  & 0.040 \\
                          &  $HR_{0\%}-HR_{100\%}$ & 31.80 & 0.90 & 19.00 & -0.07 & 12.40 & 0.038 \\
Case-IV & $IB_{0\%}-IB_{20\%}$  & 22.25 & 0.72 & 13.00 & -0.64 & 13.21 & 0.049 \\
                          & $IB_{0\%}-IB_{40\%}$ & 22.10 & 0.73 & 13.00 & -0.66 & 3.47  & 0.048 \\
                          & $IB_{0\%}-IB_{60\%}$  & 19.35 & 0.74 & 11.00 & -0.70 & 3.51  & 0.047 \\
                          & $IB_{0\%}-IB_{80\%}$ & 21.20 & 0.71 & 13.00 & -0.69 & 16.22 & 0.045 \\
                          & $IB_{0\%}-IB_{100\%}$ & 24.60 & 0.69 & 15.00 & -0.58 & 14.97 & 0.045 \\
\bottomrule
\end{tabular}

AD - Average Degree,
CC - Clustering Coefficient,
MD - Minimum Degree,\\
AC - Assortative Coefficient,
PLF - Power Law Fit,
PT - Percolation Threshold\\

\label{tab:percolation_main}
\end{table}

\subsubsection{Discussion for Case-III:}
For Case-III, extreme weather conditions are modelled considering the increase of resistance in overhead lines. From Table \ref{tab:percolation_main}, we observe average degree is high compared to other cases, implying the system is relatively highly clustered. The clustering coefficients are $\approx 0.90$ hence the network connections are effectively strong. On the other hand, assortative coefficients are almost zero for all the networks except for the network having 60\% of PV incorporated thus networks are non-assortative. A network with a given assortativity comprises its nodes that contribute to the assortativity characteristics \cite{nold}. Power law fitting values are high for all the networks which suggest that the network disobeys the power law except when incorporated with 60\% of PV in this case. From this case study, it is observed that for the only condition of having 60\% PV incorporated it may able to deal with the changes as the network is disassortative thus strong nodes connect well with weak nodes. Also, this is the only network combination following a power law. This observation from our analysis brings out a critical inference in system design for optimal PV incorporation for this topology/configuration.

\begin{figure}
  \centering
  \includegraphics[width=4.8in]{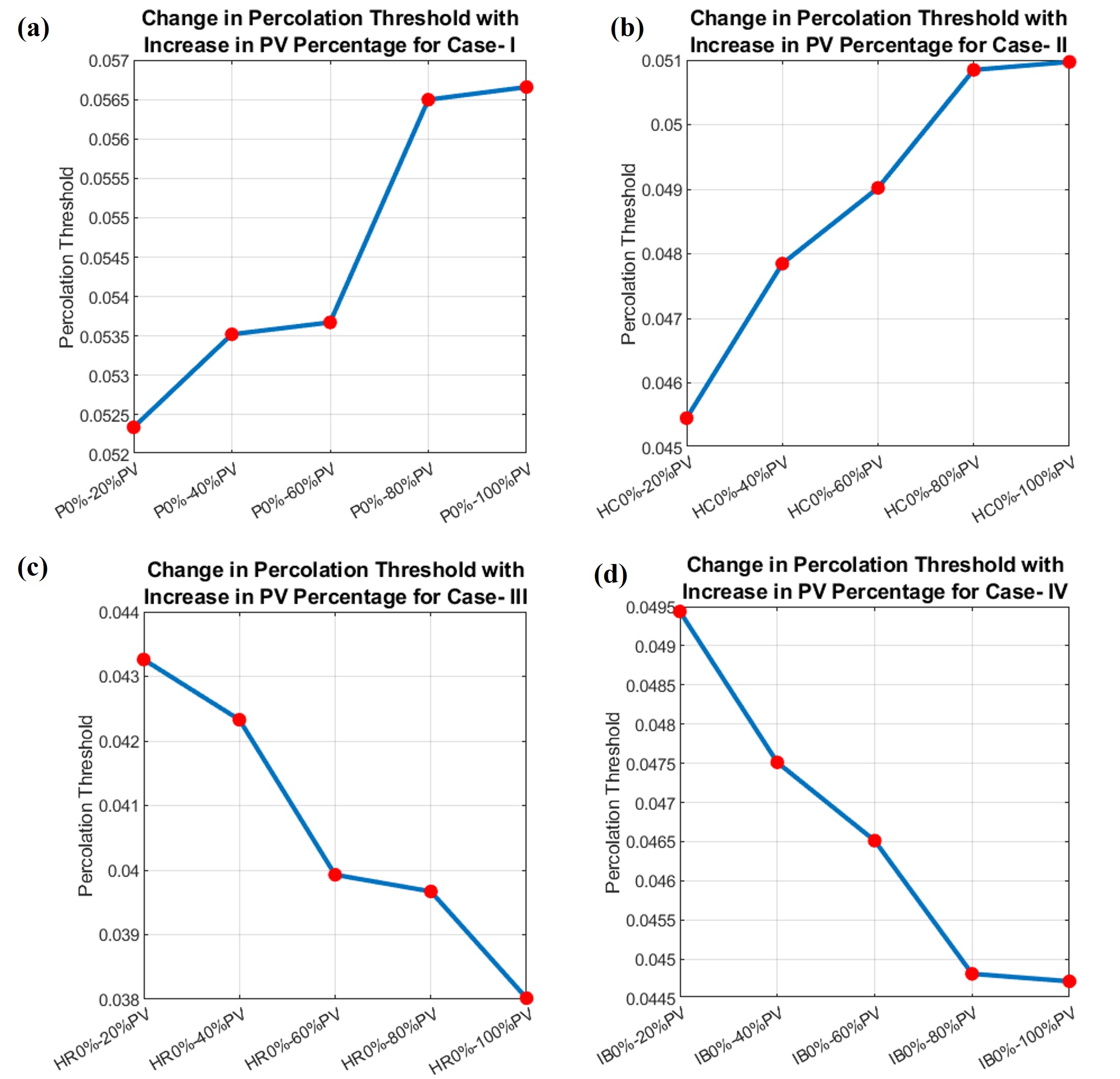}\\
  \caption{Change in percolation threshold with an increase in PV percentage for all the Cases, (a) system in the steady state becoming more resilient with an increase in PV as percolation threshold is increasing, (b) system under high loading also becomes more resilient with an increase in PV as percolation threshold is increasing, (c) in this case the system has high probability to become non-resilient with an increase in PV as percolation threshold is decreasing, (d) here system may sustain imbalance conditions with a lower percentage of PV but with PV percentage increase, the system could become non-resilient as percolation threshold is decreasing.}
  \label{fig:percolation}
\end{figure}

\subsubsection{Discussion for Case-IV:}
System imbalance is a major issue that distribution systems face. With the incorporation of PV, non-linearity is getting introduced in the system which interrupts the system’s desired operation and hinders its performance. The unequal distribution of loads between the three phases of the system causes the flow of unbalanced currents hence the line losses in the system and produces unbalanced voltage drops on the electrical lines. resilience is a good alternative to check system behaviour in such conditions. From Table \ref{tab:percolation_main}, we could observe the clustering coefficients are $\approx 0.72$ which suggests the system is strongly connected but less efficiently connected as in Case-I and Case-II. Assortative coefficients are towards the negative side thus the networks are disassortative where weak meter nodes are strongly connected with strong meter nodes and when conditions become unfavourable strong nodes could supply the weak nodes to avoid any interruption and maintain continuity. Power law fitting values are high for all the networks which suggests the network disobeys power law except when incorporated with 40\% and 60\% of PV in this case.

\subsection{Discussion for all considered cases}

\begin{figure}
  \centering
  \includegraphics[width=4.8in]{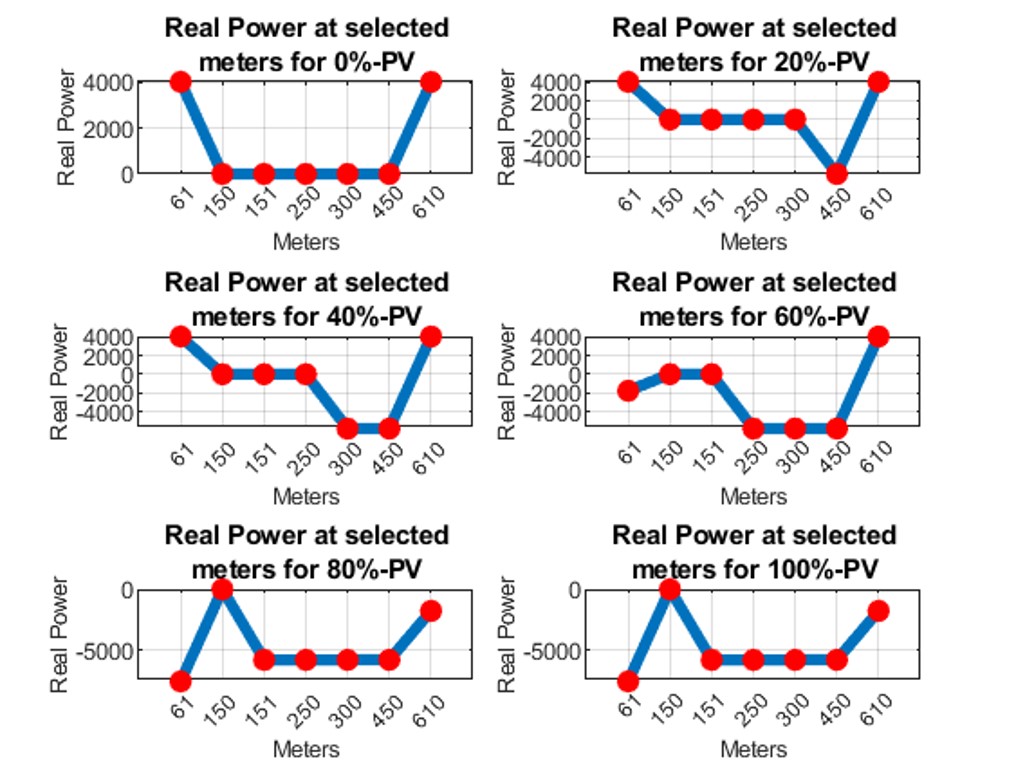}
  \caption{Real power at these selected meter nodes with an increase in PV percentage in the system eventually becomes the source node.}
  \label{fig:real}
\end{figure}

\begin{figure}
  \centering
  \includegraphics[width=3.6in]{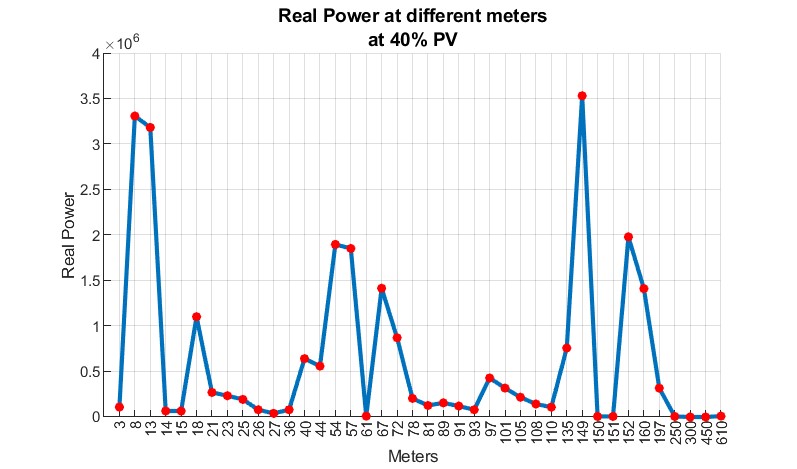}
  \caption{Real power in kW at all meter nodes when incorporated with 40\% PV in the system.}
  \label{fig:real_1}
\end{figure}

For all the cases, by computing complex network parameters including the percolation threshold, we could observe from Table \ref{tab:percolation_main} that for case-I and case-II, the percolation threshold increases with the incorporation of PV panels in the system which infers the system is becoming more resilient and could easily host PV panels. From Figure \ref{fig:percolation}(a) and \ref{fig:percolation}(b), we could infer that the IEEE 123 node test feeder is capable of maintaining resilience in normal loading as well as in high loading when incorporated with PV panels. On the other hand, for Case-III and Case-IV when the PV percentage increases, the percolation threshold is decreasing which suggests that the system may more likely transit from resilient to non-resilient. From Figure \ref{fig:percolation}(c) and \ref{fig:percolation}(d) we observe that when IEEE 123 node test feeder is loaded with resistive load and dwelled with an imbalance in loading, with PV percentage increase, the system may not cope up with changes and hence becomes non-resilient. Hence we could infer that due to the stochastic nature of DER under extreme conditions, the systems’ resilience got hampered and requires a preventive measure to avoid possible grid failure.
When PV is introduced in the system, some nodes become source nodes among the selected ones used for our analysis. These nodes are 61, 151, 250, 300, 450 and 610. It is worth emphasizing that node numbers 151, 250, 300 and 450 in their network combination have the maximum degree and hence could be considered as Hub Nodes in the network. We demonstrated the variations in power in these selected nodes with PV incorporation as shown in Figure \ref{fig:real}. For 0\% PV, real power consumed at all these meter nodes is 0 except for 61 and 610 nodes. For 20\% PV, except node 450 all nodes consumed power. Here meter node 450 continued as the source node. With 40\% PV in the system, meter nodes 300 and 450 becomes source node while nodes 61 and 610 becomes load. For 60\%, 80\% and 100\% of PV incorporation, meter nodes 61, 250, 300 and 450 becomes source node and started supplying power to other loads in the system. Similarly, meter node 151 becomes a source node with a PV percentage increased by more than 80\%. Figure \ref{fig:real_1} demonstrates the system’s real power variation at all the meter nodes when 40\% of PV is introduced. As observed, at all other nodes for this PV percentage, real power is following a similar pattern except for these selected nodes. 

\subsection{Detailed analysis and discussion for system incorporated with 40\% PV}

In this section, we discussed in detail the system behaviour with 40\% of meter nodes incorporated with PV panels to analyze by considering different combinations of PV placement. The combinations of PV placement are detailed in Table \ref{tab:Combo}. Different preselected combinations for placing PV panels at 16 meter nodes are considered for analysis to understand system behaviour when PVs are placed in combinations like cluster form (C2-C5), far-distance from each other (C13 and C14), sections controlled by switches (C7-C12), randomly scattered in the system (C1 and C6) and centrally placed combination (C15).

In Table \ref{tab:result40}, computed network parameters are tabulated for all combinations where we observe the combination C8 is at a critical state as the percolation threshold is among the lowest hence the transition from resilient to the non-resilient system is more likely to occur and the system has a higher probability to break down easily. The average degree for C8 is 33.85, which implies the system is densely connected whereas assortativity is comparatively towards the positive side hence suggesting strong nodes trying to connect with strong nodes and maintain weak relationships with weak nodes. This network about the combination C8 has a higher probability of breaking down. Similarly, as observed C12 and C15 are the next probable non-resilient system topology. From Table \ref{tab:Combo}, it is worth emphasizing that these critical combinations (i.e., C8, C12, and C15) have a PV placed at meter node 610 in common, which we may infer as the critical node for the placement of PV. 

\begin{table}
\caption{Network Parameters for different combinations under incorporation of 40\% PV}
\centering
\begin{tabular}{ccccccc}
\toprule
Graph  &	\multicolumn{1}{>{}p{1cm}}{AD} &	\multicolumn{1}{>{}p{1cm}}{CC} &	\multicolumn{1}{>{}p{1cm}}{MD}	& \multicolumn{1}{>{}p{1cm}}{AC}	& \multicolumn{1}{>{}p{1cm}}{PLF} & \multicolumn{1}{>{}p{1cm}}{PT} \\
\midrule
$P_{40\%}$-C1	& 17.750	& 0.849	& 10.000 &	-0.697 &	3.746 &	0.054 \\
$P_{40\%}$-C2	& 17.500	& 0.849	& 9.000	& -0.695	& 3.204	& 0.054 \\
$P_{40\%}$-C3	& 17.300	& 0.848	& 9.000	& -0.687	& 3.266	& 0.055 \\
$P_{40\%}$-C4	& 17.750	& 0.849	& 10.000	& -0.698	& 3.747	& 0.054 \\
$P_{40\%}$-C5	& 17.350	& 0.848	 & 9.000	& -0.686	& 3.250	& 0.058 \\
$P_{40\%}$-C6	& 17.950	& 0.848	& 10.000	& -0.698	& 2.248	& 0.050 \\
$P_{40\%}$-C7	& 18.200	& 0.817	& 9.000	& -0.715	& 2.979	& 0.049 \\
$P_{40\%}$-C8	& 33.850	& 0.851	& 24.000	& -0.158	& 35.218	& 0.034 \\
$P_{40\%}$-C9	& 17.700	& 0.849	& 10.000	& -0.700	& 3.756	& 0.051 \\
$P_{40\%}$-C10	& 18.500	& 0.824	& 10.000	& -0.718	& 3.405	& 0.051 \\
$P_{40\%}$-C11	& 18.400	& 0.809	& 10.000	& -0.717	& 3.444	& 0.054 \\
$P_{40\%}$-C12	& 32.300	& 0.809	& 22.000	& -0.109	& 21.773	& 0.036 \\
$P_{40\%}$-C13	& 18.250	& 0.817	& 10.000	& -0.715	& 3.495	& 0.052 \\
$P_{40\%}$-C14	& 17.950	& 0.824	& 10.000	& -0.699	& 2.248	& 0.050 \\
$P_{40\%}$-C15	& 31.900	& 0.795	& 24.000	& -0.212	& 26.223	& 0.039 \\
\bottomrule
\label{tab:result40}
\end{tabular}

AD - Average Degree,
CC - Clustering Coefficient,
MD - Minimum Degree,\\
AC - Assortative Coefficient,
PLF - Power Law Fit,
PT - Percolation Threshold\\

\end{table}

\begin{figure}
  \centering
  \includegraphics[width=3.0in]{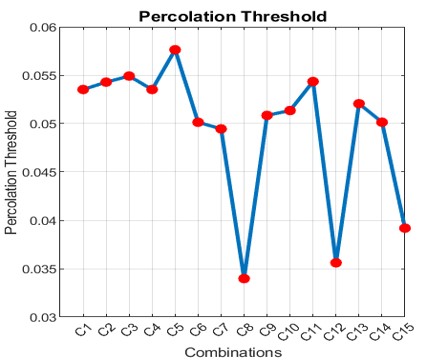}
  \caption{Percolation threshold for different combinations showing C5 as the optimal allocation topology for placing 40\% of PV in the system and the combinations C8, C12, and C15 make the system non-resilient.}
  \label{fig:percolation_40}
\end{figure}

We could also observe from Table \ref{tab:result40} that C5 is the optimal topology for placing the PV among all considered combinations as it has the highest percolation threshold among all ensuring the system will remain resilient. Figure \ref{fig:percolation_40} demonstrates a complete view of the percolation threshold variation for all different combinations. Thus, we could conclude from our analysis that this framework may be used to provide a multi-dimensional solution, as it identifies the critical nodes and also the optimal allocation topology for incorporating PV while maintaining system resilience. 

\subsection{Feature selection among network parameters}

\begin{figure}
  \centering
  \includegraphics[width=4.2in]{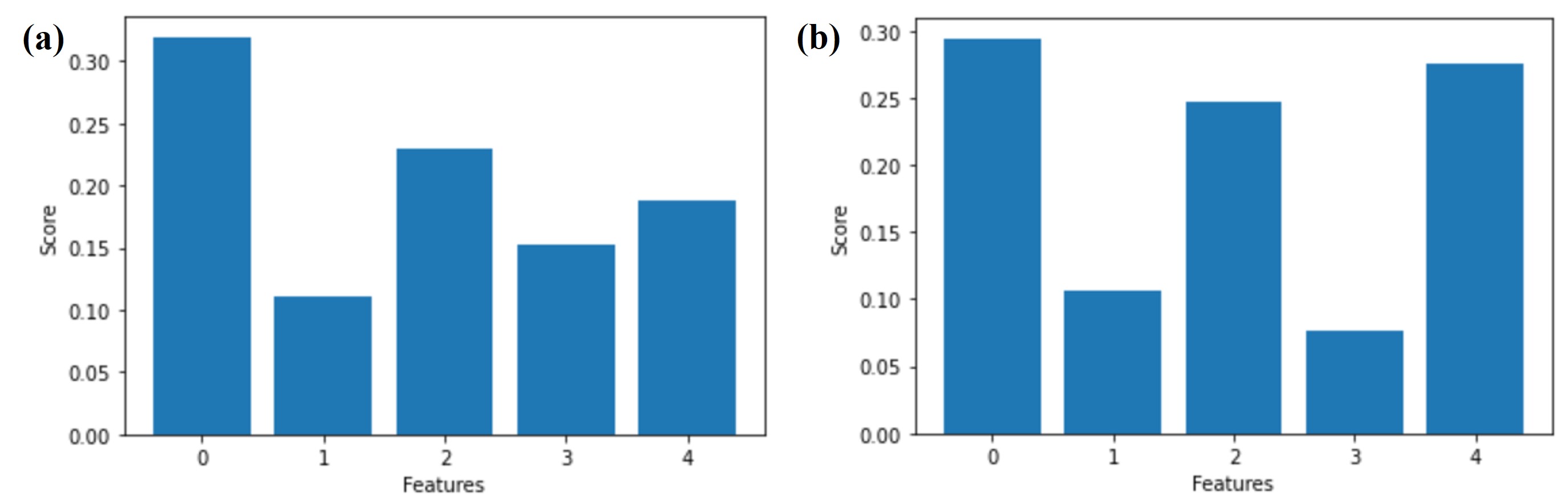}
  \caption{Bar plot depicting scores for complex network features with its importance on percolation threshold from parameters average degree, minimum degree and power law value.}
  \label{fig:score}
\end{figure}

Further to identify the network parameters that are highly sensitive to system resilience, feature selection and feature importance methodology is employed using a random forest regressor. From Figure \ref{fig:score} we observe that average degree, minimum degree and power law exponents are highly sensitive to the percolation threshold. When the average degree, minimum degree and power law exponent are high then the percolation threshold drops and vice versa. Thus, we could infer that these network parameters are the aftermath of system resilience. On the other hand, clustering and assortativity coefficients even though are important parameters of these networks but are less significant features in driving the percolation threshold as reported in Table \ref{tab:score1} and Table \ref{tab:score2}. 
           
\begin{table}
\caption{Ranking Order of features for different cases plotted in Figure \ref{fig:score}(a)}
\centering
\begin{tabular}{lll}
\toprule
Features                             & Score & Rank \\
\midrule
Average Degree – Feature -0          & 0.320 & 1    \\
Clustering Coefficient- Feature -1   & 0.111 & 5    \\
Minimum Degree- Feature -2           & 0.229 & 2    \\
Assortativity Coefficient- Feature-3 & 0.153 & 4    \\
Power Law Exponent- Feature -4       & 0.188 & 3   \\
\bottomrule
\label{tab:score1}
\end{tabular}
\end{table}

\begin{table}
\caption{Ranking Order of features for different cases plotted in Figure \ref{fig:score}(b)}
\centering
\begin{tabular}{lll}
\toprule
Features                             & Score & Rank \\
\midrule
Average Degree – Feature -0          & 0.294 & 1    \\
Clustering Coefficient- Feature -1   & 0.106 & 4    \\
Minimum Degree- Feature -2           & 0.247 & 3    \\
Assortativity Coefficient- Feature-3 & 0.077 & 5    \\
Power Law Exponent- Feature -4       & 0.273 & 2   \\
\bottomrule
\label{tab:score2}
\end{tabular}
\end{table}

Similarly, we also performed the feature selection for different combinations as mentioned in Table \ref{tab:Combo}, which had similar observations as shown in Table \ref{tab:score1} and Figure \ref{fig:score}(b). From these results, we observe that the average degree and power law exponent parameters derived from the obtained complex network equivalence of the analyzed electrical system be suitable for anticipating in case the system is prone to power interruption threats.

\section{Conclusion}
\label{section:conclusion}

To understand the resilience of an EDS with incorporated DER, we propose a hybrid data-driven methodology using complex networks and machine learning when the system undergoes undesirable conditions. This proposed method scrutinized the system resilience by rendering the real power among electrical loads and associating it with correlated networks for analysis. Further, the proposed methodology was found to be an efficient technique to check the system’s resilience with the use of various network parameters. With a gradual increase in PV incorporation in the IEEE 123 node test feeder, we identified the PV hosting capacity while maintaining the system’s resilience. The analysis shows that in normal conditions with 100\% PV incorporation, the system becomes more resilient and self-sufficient in generating the required power for the loads. For undesirable conditions such as an increase in load, increase in line resistance and imbalance in loading with PV incorporation, we observe different changing behaviour of the system from the computed network parameters that affect the system resilience.

Our proposed methodology is found to be effective in identifying the optimal allocation topology for PV in the system while maintaining its resilience. The method also identifies critical nodes of the system from the analysis that may not be suitable for placing PV as the system transit to non-resilience. We also demonstrate the effectiveness of various computed network parameters obtained from a highly correlated and dense network as it effectively identifies the phase transition of the EDS from resilient to non-resilient under several considered circumstances when introduced with DER. Furthermore, feature selection uses a random forest regressor for anticipating threats in the system.

In this study, we have evaluated the resilience of the EDS when integrated with DERs and proposed a quantifiable measure for assessment. Our method considers various operating conditions, capturing dynamic variations by calculating resilience based on the incoming real power. When the proposed method is implemented in a real-time system it effectively measures the system's resilience using the percolation threshold as a data-driven feature parameter. However, in our current study, we wish to acknowledge that the simulated model for generating the data is relied on a conservative rating assumption without incorporating the DTR system, hence it may undermine the actual line ratings. In future studies, we plan to integrate the DTR system along with DERs and compute more accurate and reliable values of percolation threshold for flexible ratings of distribution lines. Hence, it helps in optimizing infrastructure utilization through dynamic adjustments based on real-time conditions. We also plan to explore in future, the impact of peer-to-peer trading scenarios on EDS to evaluate the system's ability to adapt to changing energy dynamics which may enhance our understanding of the system's resilience and reliability.

\section*{Nomenclature}

\noindent The following abbreviations are used in this manuscript:

\noindent EDS - Electrical Distribution Systems\\
    DER - Distributed Energy Resources\\
    IEEE - Institute of Electrical and Electronics Engineers\\
    EDN - Electrical Distribution Networks\\
    PV - Photovoltaic\\
    AD - Average Degree\\
    CC - Clustering Coefficient\\
    MD - Minimum Degree\\
    AC - Assortative Coefficient\\
    PLF - Power Law Fit\\
    PT - Percolation Threshold\\
    DTR - Dynamic Thermal Rating

\section*{Acknowledgement}
The Authors would like to thank Mr. Siddharth Patwardhan, Dr. P. Manimaran for their useful discussion on various complex network topics and their significance, and Dr. Alok Kumar Bharati for the discussion on GridLAB-D.

\section*{CRediT authorship contribution statement}
Divyanshi Dwivedi participated in idea generation, implemented, and evaluated the ideas, performed data analysis, and wrote the manuscript. Pradeep Kumar Yemula contributed to the discussion, guidance, and review of the manuscript. Mayukha Pal conceived the idea and conceptualized it, prepared the analysis methodology, mentored in results analysis and discussion, project guidance, and wrote and reviewed the manuscript.

\section*{Declaration of Competing Interest}
The authors declare that they have no known competing financial interests or personal relationships that could have appeared to influence the work reported in this paper.

\bibliographystyle{elsarticle-num}

\bibliography{main}
\end{document}